\documentstyle[pra,aps,epsf,twocolumn,floats,times]{revtex}

\title{High efficiency entangled photon pair collection in type II parametric fluorescence}
\author{Christian Kurtsiefer$^1$, Markus Oberparleiter$^1$ and Harald Weinfurter$^{1,2}$}
  \address{$^1$ Sektion Physik, Ludwig-Maximilians-Universit\"at, D-80797 M\"{u}nchen, Germany}
 \address{$^2$ Max-Planck-Insitut f\"{u}r Quantenoptik, D-85748 Garching, Germany}
\date{December 19, 2000}
\draft
\begin{document}
\maketitle

\begin{abstract}
We report on a method for optimizing the collection of entangled photon pairs
in type-II parametric fluorescence. With this technique, we detected
360000 polarization-entangled photon pairs per second in the near-IR
region in single-mode optical fibers. The entanglement of the photon pairs 
was verified by measuring polarization correlations in different bases of at
least 96\%.
\end{abstract}

%
% 42.65.Yj  Optical parametric oscillators and amplifiers
% 42.50.Ar Photon statistics and coherence theory
% 42.50.Dv Nonclassical field states; squeezed, 
%         antibunched, and sub-Poissonian states; 
%         operational definitions of the phase of the 
%         field; phase measurements


Experiments with
entangled photon pairs have opened a whole field of research: Created
initially by electron-positron annihilation and later in atomic cascade
decays, entangled photons allowed a distinctive comparison of various concepts
of Quantum mechanics\cite{clauser,aspect}. 
More recently, parametric fluorescence (down-conversion) in nonlinear optical
crystals as the source of entangled photon pairs\cite{mandel85,rarity,kwiat95}
lead to a dramatic increase in count rates.
This enabled a variety of experiments on the foundations of quantum
mechanics\cite{mandel99,zeili99,chiao96} and the experimental realization of
new concepts in quantum information\cite{boutele,bouGHZ,harald}. In spite of
that success, most of the experiments and potential applications still suffer
from the low yield of the fluorescence process. 
It was thus the motivation for the work presented here to optimize collection
efficiency and thereby the available rate of polarization-entangled photon
pairs from parametric down conversion.

In type-II parametric fluorescence, a pump photon with energy 
$\hbar\omega_p$ is converted in a nonlinear optical crystal into two
orthogonally polarized photons, signal and idler, obeying energy and momentum
conservation.
For a fixed pump photon momentum and a given idler frequency $\omega_i$,
there is a one-dimensional manifold of emission directions for the idler
photon, and a corresponding one for the signal photon with a frequency
$\omega_s$ with $\omega_p=\omega_i+\omega_s$, forming two emission cones. To
generate polarization entangled photon pairs, the orientation of the nonlinear
crystal is chosen such that the two cones intersect. This intersection defines
two directions along which the polarization of each emitted photon is
undefined, but perfectly anti-correlated with the polarization of the other
one. Provided complete indistinguishability of which photon belongs to 
which emission cone, a polarization-entangled pair of photons is
obtained\cite{kwiat95}.

In most of the experiments performed, the photons have been collected into
spatial modes defined by apertures, and the spectrum of the
collected light has been defined with optical filters to a given bandwidth
$\Delta\lambda$ around the wavelengths $\lambda_i$ and $\lambda_s$ of signal
and idler photons. 
Yet, in many interferometric experiments requiring a well-defined spatial mode
or for transport over larger distances it is desirable to couple the light from
the parametric down conversion into single mode optical fibers. With specific
design criteria in an experiment  
testing Bell inequalities with space-like separated observers\cite{weihs},
typical pair rates of 13000~s$^{-1}$ for $\Delta\lambda_{\rm FWHM}=3$~nm at a
pump power of 400~mW have been achieved (crystal thickness
3~mm)\cite{gwdiss}. The collection efficiency there was
already an order of magnitude higher than in the initial experiments using
type-II down conversion sources\cite{kwiat95}. 
 Different techniques have been implemented since then to increase the number
of entangled photon pairs emitted in a single spatial mode, including the use
of two
Type-I conversion crystals\cite{kwiat00}, focusing the pump
beam\cite{monken1,monken2}, or using resonant enhancement\cite{ou99,markuso}.
Quite recently, several groups have succeeded in designing photon pair sources
with confining waveguides in periodically poled crystal
structures\cite{ppstuff,pp2}, showing unprecedented efficiencies in the
generation of correlated photon pairs. However, to our knowledge, no entangled 
photons have been observed with this technique yet.

Here, we present a simple way to optimize the collection efficiency, using the
fixed relation between emission direction and wavelength for a fixed pump
frequency.
The idea is to match the angular distribution of the parametric fluorescence
light for a given spectral bandwidth to the angular width of the spatial mode 
collected into a single mode optical fiber.
Applying this design criterium, which also can be combined with the new
methods mentioned above, we achieved unprecedented coupling efficiencies and
photon pair detection rates.

\begin{figure}
\centerline{\epsfxsize\linewidth\epsffile{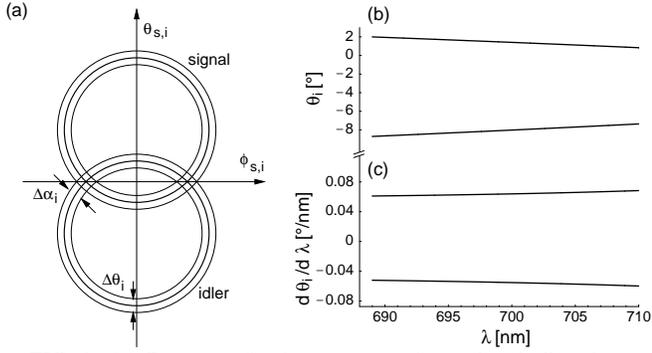}}
\caption{\label{theobild}(a) Geometry for the emission of signal and idler
  photons for a finite bandwidth. (b) Emission angle $\theta_i$ of idler
  relative to the pump beam for $\Theta_p=49.7^\circ$ and $\phi_i=0^\circ$. (c)
  Derivative $d\theta_i/d\lambda$ to estimate the angular width
  $\Delta\alpha_i$ to be collected.} 
\end{figure}

To optimize collection efficiency, we consider the wavelength dependency of the
opening angles of the emission cones. For a given spectral width
$\Delta\lambda_i=\Delta\lambda_s$, the signal and idler light emitted along
the intersection directions is dispersed over an angular width
$\Delta\alpha_i$ and $\Delta\alpha_s$, respectively.
We use the approximate rotational symmetry of the emission cones and obtain
\begin{equation}
\label{formel1}\Delta\alpha_i=\Delta\alpha_s\approx\Delta\theta_s=\Delta\theta_i={d\theta_i\over
  d\lambda_i}\Delta\lambda_i\quad,
\end{equation}
where $\theta_s$ and $\theta_i$ are the emission angles between the pump
direction and signal and idler light, respectively, in a plane containing the
optical axis of the (uniaxial) crystal (see the rings of intersection with a
plane normal to the pump beam in figure \ref{theobild}(a)).
This expression can be obtained from energy and momentum conservation
in a closed form, although the numerical solution is faster.
The pump light is considered as  a plane wave propagating with an
angle $\theta_p$ with respect to the optical axis of the nonlinear crystal.
A more detailed discussion of the relation between spectral and angular
distribution of the down-converted light can be found in\cite{rubin}.

\begin{figure}
\centerline{\epsfxsize\linewidth\epsffile{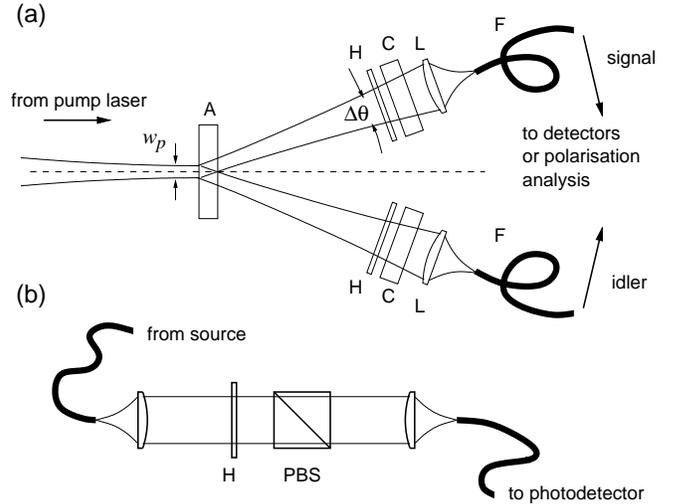}}
\caption{\label{basicsetup}Basic setup: (a) A UV pump laser beam is focused
  into 
  a BBO-crystal A to a waist $w_p$; light emitted by parametric fluorescence
  is mapped to the receiving modes of single mode optical fibers $F$
  with lenses $L$. A combination of half-wave-plates $H$ and
  additional BBO-crystals $C$ are used to compensate for walk-off in the first
  crystal. (b) Polarization analysis in each arm with an additional half wave
  plate $H$ selecting the measurement basis and a polarizing
  beam splitter $PBS$. }
\end{figure}

For an appropriate choice of the crystal orientation, the two directional
manifolds 
of signal and idler light intersect perpendicularly. Then, for a Gaussian
spectral distribution of the light to be collected, the corresponding angular
distributions are also Gaussian with characteristic widths
$\Delta\alpha_s=\Delta\alpha_i$, and have rotational symmetry around the
intersection directions. 
We now define Gaussian target modes aligned with the intersection
directions of the emission cones with a divergence
$\theta_D=\Delta\alpha_{s,i}$,  which can be mapped optically to the
collecting fibers.
The beam waist of these Gaussian modes is
given by $w_0=\lambda/(\pi\theta_D)$. As the mode matching for the
parametric down conversion is described in a plane wave basis, we locate the
waist of this Gaussian mode in the conversion crystal, as sketched in figure
\ref{basicsetup}(a). For our configurations, the crystal is always shorter
than the Rayleigh length $z_r=\pi w_0^2/\lambda$ of the corresponding modes,
thus we neglect possible effects due to wavefront curvature in the conversion
crystal.

Having chosen the target modes, we only collect photons being created in a
region of overlap between target modes and pump field. Therefore, the pump
field can be restricted to a region where the target modes have a significant
field strength. To maximize overlap with the Gaussian pump field, we
choose its waist $w_p$ to be equal to the waist of the target modes in the
crystal.

As in previous work\cite{kwiat95}, we use an Argon ion laser in our experiment
at a wavelength of  $\lambda_p=351.1$~nm to pump a BBO nonlinear optical
crystal (thickness 2~mm),
yielding down converted photon pairs efficiently
detectable with Si avalanche diodes.

The two cones 
of signal and idler light at the degenerate
wavelength $\lambda_s=\lambda_i=702.2$~nm intersect perpendicularly for a 
pump beam orientation of $\Theta_p=49.7^\circ$ with respect to the optical
axis, resulting in an (external) angle of $\phi_{s,i}=\pm3.1^\circ$ with
respect to the pump beam (figure \ref{theobild}(a)).

Figures \ref{theobild}(b,c) show both the emission angles $\theta_i, \theta_s$
and their derivatives for that configuration. The results 
are obtained  numerically, with refractive indices of the crystal
given by numerical Sellmeier equations\cite{eimerl}. The
angles include the refraction on the crystal surfaces, assuming 
the crystal faces oriented normal to the pump beam direction. At the
degeneracy wavelength of $\lambda=702.2$~nm, we obtain an angular derivative
of $|d\theta_i/d\lambda_i|=|d\theta_s/d\lambda_s|\approx0.055^\circ$/nm.

Using the asymptotic expression
\begin{equation}
I(\theta)\propto exp(-2\theta^2/\theta_D^2)
\end{equation}
for the intensity distribution $I(\theta)$ in a Gaussian target mode, and
aiming for a spectral width of 
$\Delta\lambda_{\rm FWHM}=4$~nm for the down converted photons, one would
expect a divergence angle of 
\begin{equation}
\theta_D\approx\Delta\lambda_{\rm
  FWHM}/\sqrt{2\ln2}\times|d\theta_i/d\lambda_i|=0.186^\circ.
\end{equation}
To compensate for an additional divergence due to the angular distribution in
the pump beam, we have chosen a divergence angle $\theta_D=0.16^\circ$ for the
target 
modes, corresponding to a Gaussian beam waist of $w_0=82~\mu$m. These modes
were geometrically mapped with aspheric lenses ($f=11$~mm) to the receiving
modes of single mode optical fibers with a Gaussian waist parameter
determined to be $w_f=2.3~\mu$m.

Transverse and longitudinal walk-off in the conversion crystal were
compensated in the usual way\cite{kwiat95} with additional 
polarization rotators and BBO crystals of length 1~mm to obtain polarization
entangled photons for all wavelengths in the acceptance spectrum of the
down-converted light (figure \ref{basicsetup}(a)).
In our arrangement, we used interference filters for bandwidth selection in
front of the couplers only 
for an initial alignment; the experimental data described below were obtained
without these filters. 

\begin{figure}
\centerline{\epsfxsize\linewidth\epsffile{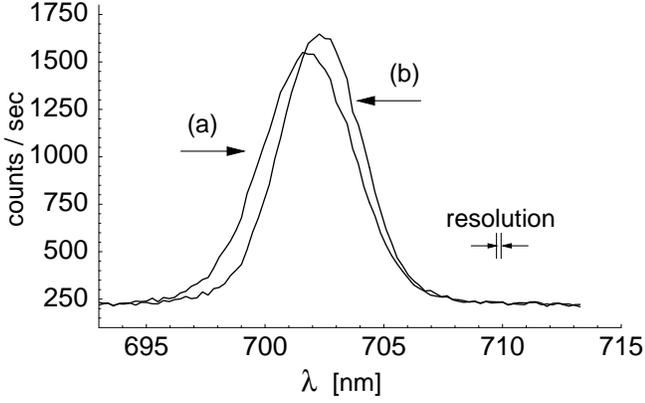}}
\caption{\label{spectra}
Optical spectra of light collected from the down conversion into the two
single mode fibers. The spectra show a FWHM of 4.60~nm (a) and 4.06~nm
(b) and are centered around the degeneracy wavelength of 702.2~nm with a
separation of 0.55~nm.  
}
\end{figure}

Figure \ref{spectra} shows the spectral distribution of the converted light
collected into the single mode fibers. The spectra in both arms exhibit an
almost Gaussian distribution with a FWHM of 4.06~nm and 4.60~nm,
respectively\cite{bemerk1}. They are centered 
around the degeneracy wavelength of 702.2~nm, with a separation of 0.55~nm
due to residual misalignment of the fiber coupler positions. The widths of the
observed spectra are in good agreement with the estimation given above.

\begin{figure}[b]
\centerline{\epsfxsize\linewidth\epsffile{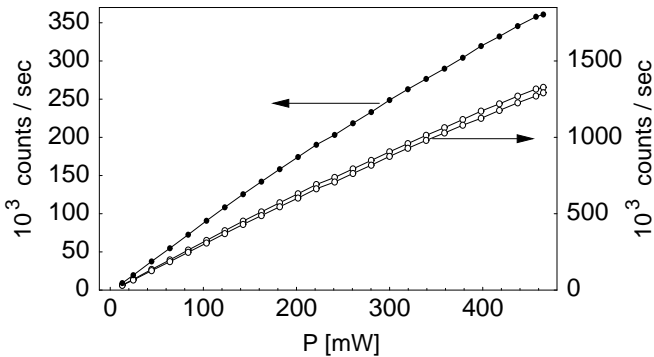}}
\caption{\label{powdep}
Coincidence count rates (left axis) and individual count rates (right axis) of
two photodetectors connected directly to the single mode fibers as a function
of UV power.
}
\end{figure}

To determine the collection efficiency, we connected the single mode optical
fibers directly to two actively quenched silicon avalanche
diodes.  Figure \ref{powdep} shows the coincidence
count rates and the count rates for the individual detection events as a
function of UV pump power. We determined a coincidence/single count ratio
(i.e., an overall efficiency) of $0.286\pm.001$ for the whole range of pump
power. The maximum coincidence count rate we observed was 360800~s$^{-1}$ at a
pump power of $P=465$~mW\cite{saettigung}. The coincidence time window was
measured to be 
$\tau_c=6.8\pm0.1$~ns, thus accidental coincidence count rates being small over
the whole range of pump power. For the low power regime, we
obtain a slope of  900 coincidence counts per second and mW for our 2~mm long
BBO crystal in the single mode optical fibers.

We like to point out that the coincidence/single ratio may significantly be
affected by transverse walk-off of the extraordinary beam: The transverse
walk-off for the passage through the down-conversion crystal is
170~$\mu$m, which has to be compared to a receiving waist of
82~$\mu$m. For experiments requiring even higher coincidence/single ratio,
e.g. in the attempt of closing the detector loophole in an EPR experiment, one
would choose a smaller optical bandwidth, causing a larger beam waist of the
target mode and thereby a less dramatic influence of the transverse walk-off.
Another possibility would be the use of cylindrical lenses.

To verify the entanglement of the photon pairs obtained with this collection
arrangement, we measured the polarization
correlation between the photons of each pair using a polarization analyzer as
shown in figure \ref{basicsetup}(b) formed by a half wave plate and a
polarizing beam splitter. The birefringence of the optical fiber was
compensated with a 'bat ear' polarization controller such that less than 1\% of
intensity was transfered from $0^\circ$ or $45^\circ$ linear polarization to
the corresponding orthogonal polarization.
After polarization analysis, the light was coupled into a multi-mode
optical fiber, and detected by the two actively quenched silicon avalanche
diodes.

\begin{figure}
\centerline{\epsfxsize\linewidth\epsffile{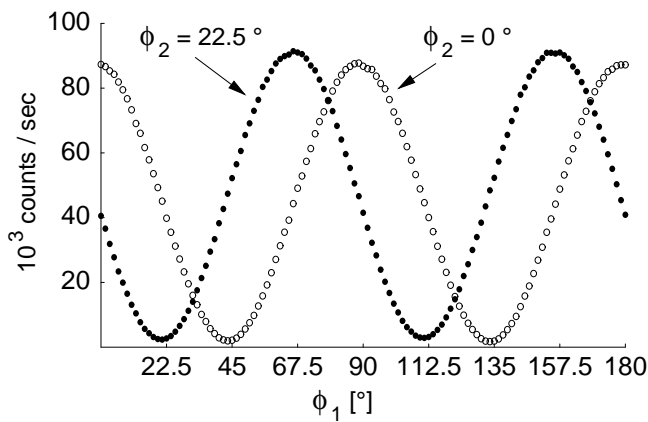}}
\caption{\label{visis}
Polarization correlation experiment: Coincidence count rates of
the polarization analysis setup for 
different orientation angles $\phi_1,\phi_2$ of the  half-wave
plates. Uncorrected polarization correlation in the (H,V) basis is 96\%, and
95\% in the $\pm45^\circ$ basis.
}
\end{figure}

The single and coincidence count rates for different orientations
$\phi_1,\phi_2$ of the half wave plates in the analyzers are shown in figure
\ref{visis} for a UV pump power of 400~mW. For a setting of $\phi_2=0^\circ$, 
corresponding to detecting horizontally polarized photons, we
observed a visibility of the polarization correlation of $96.0\pm0.1$\%. This
value was obtained from a sin/cos fit to the experimentally obtained
coincidence count rates. To correct for accidental coincidences, we subtract a
coincidence rate of $n_c=n_sn_i\tau_c(1-\eta)$, with a total detection
efficiency 
$\eta=0.214$ in the polarization analysis setup, and single count rates of
$n_s,n_i\approx420000$ counts per second. With that, we find a visibility of
$98.2\pm0.1$\%. For a setting of
$\phi_2=22.5^\circ$ corresponding of a detection of $+45^\circ$ linearly
polarized photons in arm 2, the coincidence count rate as a function of the
tilt angle $\phi_1$ is shown in figure \ref{visis}. From this measurement,
we obtain a bare visibility of $94.5\pm0.1$\% for the polarization
correlation, and $96.3\pm0.1$\% after correction for accidentals.
The corrected visibilities of the polarization correlation both in a (H,V) and
a ($+45^\circ,-45^\circ$) basis are compatible with values of $98.8\pm0.2$\%
and $97.0\pm0.2$\%  we obtained at a lower pump power of $P=123$~mW, where
accidental coincidences are negligible.
Using (raw) data from the high intensity experiment for the evaluation of a
CHSH-type Bell inequality, we obtain $S=-2.6979\pm0.0034$, i.e., a violation of
204 standard deviations for a measurement time of one second per angle setting.

To summarize, we presented a technique to optimize the  collection of
entangled photon pairs created in a type-II parametric fluorescence process
using mode matching procedures. We thereby were able to increase pair
collection efficiencies of earlier experiments\cite{gwdiss} into single mode
optical fibers  by a factor of $\approx25$, supplying a strong source for
quantum information experiments having attracted attention
recently. It is particularly interesting to combine these techniques with
resonant enhancement techniques\cite{markuso}, since this leads to strong
sources for polarization-entangled photon pairs without the need for costly
large-frame lasers\cite{juergen}.

\acknowledgments
We like to acknowledge the uncomplicated help from H.~Zbinden and N.~Gisin
from the University of Geneva with actively quenched photodetectors. This
work was supported by the European Union in the FET/QuComm research project (IST-1999-10033) and the Deutsche Forschungsgemeinschaft.


\begin{references}
% aspect
\bibitem{clauser}J.F.~Clauser and A.~Shimony, Rep. Prog. Phys. {\bf 41}, 1981
  (1978).
\bibitem{aspect}A.~Aspect, P.~Grangier, and G.~Roger, Phys. Rev. Lett. {\bf
    47}, 460 (1981); {\bf 49}, 91 (1982); A.~Aspect, J.~Dalibard and R.~Roger,
  Phys. Rev. Lett. {\bf 49}, 1804 (1982).

% mandel urgestein arbeit
\bibitem{mandel85}C.K.~Hong and L.~Mandel, Phys. Rev. A {\bf 31}, 2409 (1985).

% old: Rarity & friends
\bibitem{rarity}J.G.~Rarity, P.R.~Tapster, Phys. Rev. Lett. {\bf 64}, 2495
  (1990).

% kwiat & friends; shih
\bibitem{kwiat95}P.G.~Kwiat, K.~Mattle, H.~Weinfurter, A.~Zeilinger,
  A.V.~Sergienko, Y.~Shih, Phys. Rev. Lett. {\bf 75}, 4337 (1995).

% Uebersichtsartikel:
\bibitem{mandel99}L.~Mandel, Rev. Mod. Phys. {\bf 71}, S274 (1999).
\bibitem{zeili99}A.~Zeilinger, Rev. Mod. Phys. {\bf 71}, S288 (1999).
\bibitem{chiao96}A.E.~Steinberg, P.G.~Kwiat, and R.Y.~Chiao in {\it Atomic,
    molecular and optical physics handbook (ed. G.~Drake)}, Ch.~77, 901, AIP
  press, New York (1996).

% teleport:
\bibitem{boutele}D.~Bouwmeester,~J.-W. Pan, K.~Mattle, M.~Eibl, H.~Weinfurter,
  and A.~Zeilinger, Nature {\bf 390}, 576 (1997).
% GHZ
\bibitem{bouGHZ}D.~Bouwmeester, J.-W.~Pan, M.~Daniell, H.~Weinfurter, and
  A.~Zeilinger, Phys. Rev. Lett. {\bf 82}, 1345 (1999).

% allgemeines quantum communication review
\bibitem{harald}H.~Weinfurter, Adv. At. Mol. Opt. Phys. {\bf 42}, 489 (2000).


% long distance EPR
\bibitem{weihs}G.~Weihs, T.~Jennewein, C.~Simon, H.~Weinfurter, and
  A.~Zeilinger, Phys. Rev. Lett. {\bf 81}, 5039 (1998).
% Gregors Diss
\bibitem{gwdiss}G.~Weihs, PhD thesis, Universty of Vienna, 1999.

% more kwiat: type-1 stuff
\bibitem{kwiat00}P.G.~Kwiat, E.~Waks, A.G.~White, I.~Appelbaum, P.H.~Eberhard,
    Phys. Rev. A {\bf 60}, R773 (1999).

% brasilianer Zugang
% optimizing....for bell measurement
\bibitem{monken1}C.H.~Monken, P.H.~Suoto Ribeiro, and S.~P\'adua,
  Phys. Rev. A {\bf 57}, R2267 (1998).
% transfer of angular spectrum...
\bibitem{monken2}C.H.~Monken, P.H.~Suoto Riberio, and S.~P\'adua, Phys. Rev. A
  {\bf 57}, 3123 (1998).

% resonator work
\bibitem{ou99}Z.Y.~Ou and Y.J.~Lu, Phys. Rev. Lett. {\bf 83}, 2556 (1999).
\bibitem{markuso} M.~Oberparleiter and H.~Weinfurter, Opt. Com. {\bf}183,
  133-137 (2000).

% neues ppln-Zeug
\bibitem{ppstuff}
S.~Tanzilli, H.~de~Riedmatten, W.~Tittel, H.~Zbinden, P.~Baldi, M.~de~Micheli,
D.B.~Ostrowski and N.~Gisin, to appear in Electron. Lett.

% Spektrometerbemerkung
\bibitem{pp2} K.~Sanaka, K.~Kawahara, and T.~Kuga,  quant-ph/0012028.

% saettigungsbemerkung
\bibitem{saettigung} To avoid saturation effects in the counter electronics,
  we used a fast $\div16$ pre-scaler for both single and coincidence counts.

% rubinstuff
\bibitem{rubin}M.H.~Rubin, Phys. Rev. A {\bf 54}, 5349 (1996).

% good Sellmeier equations for BBO
\bibitem{eimerl}D.~Eimerl, L.~Davis, S.~Velsko, E.K.~Graham, and A.~Zalkin,
  J. Appl. Phys. {\bf 62}, 1968 (1987).

\bibitem{bemerk1}The resolution of the spectrometer was measured with a
  HeNe reference laser to be $\Delta\lambda_{\rm FWHM}=0.2$~nm.

%juergens diplomarbeit
\bibitem{juergen}J.~Volz, Diploma thesis, University of Munich, (2000).

\end{references}
\end{document}